\documentclass[runningheads]{llncs}
\usepackage{setup}

\title{On the effect of social norms on performance in teams with distributed decision makers}
\titlerunning{Effect of social norms on team performance}

\author{Ravshanbek Khodzhimatov\inst{1}\orcidID{0000-0002-2761-2029} \and
Stephan Leitner \inst{2}\orcidID{0000-0001-6790-4651} \and
Friederike Wall\inst{2}\orcidID{0000-0001-8001-8558}}

\authorrunning{R. Khodzhimatov et al.}

\institute{Digital Age Research Center, University of Klagenfurt, 9020 Klagenfurt, Austria
\email{ravshanbek.khodzhimatov@aau.at}\\
\and
Department of Management Control and Strategic Management, University of Klagenfurt, 9020 Klagenfurt, Austria\\
\email{\{stephan.leitner, friederike.wall\}@aau.at}}

\begin{document}
\maketitle 
\begin{abstract}
Social norms are rules and standards of expected behavior that emerge in societies as a result of information exchange between agents. This paper studies the effects of emergent social norms on the performance of teams. We use the $N\!K$-framework to build an agent-based model, in which agents work on a set of interdependent tasks and exchange information regarding their past behavior with their peers. Social norms emerge from these interactions. We find that social norms come at a cost for the overall performance, unless tasks assigned to the team members are highly correlated, and the effect is stronger when agents share information regarding more tasks, but is unchanged when agents communicate with more peers. Finally, we find that the established finding that the team-based incentive schemes improve performance for highly complex tasks still holds in presence of social norms.

\keywords{Agent-based modeling and simulation  \and $NK$-framework \and emergence \and socially accepted behavior}

\end{abstract}
\section{Introduction}
\label{sec:intro}
One of the main goals of team managers is to design framework conditions that allow teams to achieve a high performance. This, amongst others, includes the allocation of tasks and choice of the means of behavioral control, such as incentive schemes. Whenever people collaborate, however, there might be emergent social dynamics that probably interfere with an organization's or the managers' decisions on these framework conditions. In this paper, we address two questions related to emergent social norms and means of behavioral control: first, \textit{how do emergent social norms affect a team's performance} and, second, \textit{how is the efficiency of behavioral control mechanisms affected by the presence of social norms}. These questions are becoming even more apparent now, with continuous developments in digitalization and the shift of teams towards digital communication channels in time of a global pandemic.

In our research, we particularly address \textit{descriptive} social norms, which Cialdini et al. \cite{cialdini90} defined as the frames of reference that emerge when individuals observe and adopt the \textit{actual behavior} of their peers, as opposed to normative moral claims that prescribe certain predefined actions to individuals. Priebe et al. \cite{priebe11} summarized the concept descriptive social norms as: \textit{``When in Rome, do as the Romans do''}. An example of descriptive social norms is choosing \LaTeX\ instead of PowerPoint to prepare presentations in the absence of any external requirement only because teammates do so. 

The idea of descriptive social norms is not new but has been formalized, among others, by Cialdini et al. \cite{cialdini90}. They argued that descriptive norms serve as ``decisional shortcuts'': \textit{``If everyone is doing it, it must be a sensible thing to do''}. Pryor et al. \cite{pryor19} argued that another reason why people conform to descriptive norms is to diffuse responsibility for a risky action by imitating what they perceive is the status quo.

Since there are potentially many confounding variables, it is difficult to find an empirical evidence for whether and, if so, how individual's decisions are driven by social norms (in the above sense) \cite{wall21}. Thus, in order to investigate this potential effect, we build a stylized agent-based model, in which descriptive norms are formed as individuals share and observe the behavior of their teammates.

In the context of agent-based simulations, the idea of \textit{imitation} as a means to increase individual performance is not new. Rivkin \cite{rivkin00} used the $NK$-framework \cite{kauffman89,levinthal97} to model a firm that imitates an industry leader, and found that, in presence of complexity, the imitation does not result in an increase in performance.
In this paper we follow a different approach: we do not model agents that directly imitate other agents but we focus on agents that conform with what appears to be socially acceptable behavior (which, of course, might be affected by imitation, amongst others). In particular, we model a team that works on a complex task and the team members are assigned parts of the task, which may or may not be interdependent. As the team members communicate, they share information about their past decisions, which forms the basis for the emergence of descriptive social norms. We investigate the effect of social norms on the functioning of the incentive schemes, which are adopted to control the team members' behavior, and observe the team's overall performance for environments with different task structures.
\section{Model}
\label{sec:model}
In this section we introduce the agent-based model of a team of $P=4$ individuals facing a complex task. The task environment is based on the $NK$-framework \cite{kauffman89,levinthal97}. Agents make decisions to (a) increase their compensation (based on their performances) and (b) comply with the descriptive social norms. Sec. \ref{sec:design} introduces the task environment, Secs. \ref{sec:agents} and \ref{sec:social} characterize the agents and describe how social norms emerge, respectively. Sec. \ref{sec:discovering} describes the agents' search process, and Sec. \ref{sec:process} provides an overview of the sequence of events in the simulation.

\subsection{Task environment}
\label{sec:design}

We model an organization that faces a complex decision problem that is expressed as the vector of $M=16$ binary choices. The decision problem is divided into sub-problems which are allocated to $P=4$ agents, so that each agent faces an $N=4$ dimensional sub-problem:
\begin{equation}
    \mathbf{x} = (
        \underbrace{x_1,x_2,x_3,x_4}_{\mathbf{x}^1},
        \underbrace{x_5,x_6,x_7,x_8}_{\mathbf{x}^2},
        \underbrace{x_9,x_{10},x_{11},x_{12}}_{\mathbf{x}^3},
        \underbrace{x_{13},x_{14},x_{15},x_{16}}_{\mathbf{x}^4}
        ),
    \label{eq:tasks}
\end{equation}
where bits $x_i \in \{ 0, 1 \}$ represent single tasks.
Every task $x_i$ is associated with a uniformly distributed performance contribution $\phi(x_i) \sim U(0,1)$. The decision problem is \textit{complex} in that the performance contribution $\phi(x_i)$, might be affected not only by the decision $x_i$, but also by decisions $x_j$, where $j \neq i $. We differentiate between two types of such inter-dependencies: (a) \textit{internal}, in which interdependence exists between the tasks assigned to agent $p$, and (b) \textit{external}, in which interdependence exists between the tasks assigned to agents $p$ and $q$ for $p \neq q$. We control inter-dependencies by parameters $K,C,S$, so that every task interacts with exactly $K$ other tasks internally and $C$ tasks assigned to $S$ other agents externally \cite{kauffman91}:
\begin{equation}
    \phi (x_i) = \phi (x_i ,
        \underbrace{x_{i_1},...,x_{i_K}}_{\substack{K\textbf{ internal}\\\text{interdependencies}}}, \underbrace{x_{i_{K+1}} ,...,x_{i_{K + C \cdot S}}}_{\substack{C \cdot S \textbf{ external}\\ \text{interdependencies}}}),
    \label{eq:payoff}
\end{equation}
where $i_1, \dots , i_{K+C\cdot S}$ are distinct and not equal to $i$. We consider two cases: (i) \textit{low complexity} (only internal interdependence: $K = 2, C = S = 0$) and (ii) \textit{high complexity} (internal and external interdependence: $K=C=S=2$).\footnote{The exact choice of the coupled tasks is random with one condition: every task affects and is affected by exactly $K+C\cdot S$ other tasks} Using Eq. \ref{eq:payoff}, we generate \textit{performance landscapes}\footnote{A performance landscape is a matrix of uniform random variables that correspond to every combination of $1 + K + C \cdot S$ decisions. We generate entire landscapes to find the overall global maximum and normalize our results accordingly, to ensure comparability among different scenarios.} for all agents.

We are interested in the dynamics of the team's overall performance throughout $T=500$ time periods. At each time period $t$, agent $p$'s performance is a mean of performance contributions of tasks assigned to that agent:
\begin{equation}
    \phi_{own} (\mathbf{x}^p_t) = \frac{1}{N} \sum_{x_i \in \mathbf{x}^p_t} \phi(x_i),
    \label{eq:performance-agent}
\end{equation}
and the team's overall performance is a mean of agents' performances:
\begin{equation}
    \Phi (\mathbf{x}_{t}) = \frac{1}{P} \sum_{p=1}^{P} \phi_{own}(\mathbf{x}^p_t)~.
    \label{eq:performance-org}
\end{equation}

%\begin{figure}[!tb]
%    \centering
%    \begin{minipage}{0.4\linewidth}
%        \centering
%        \includegraphics[width=0.95\linewidth]{fig/interactions200}
%        \captionsetup{justification=centering}
%        \caption*{Low complexity\\{\small ($K=2,C=S=0$)}}
%    \end{minipage}
%    \begin{minipage}{0.4\linewidth}
%        \centering
%        \includegraphics[width=0.95\linewidth]{fig/interactions222}
%        \captionsetup{justification=centering}
%        \caption*{High complexity\\{\small ($K=C=S=2$)}}
%    \end{minipage}
%    \caption{Stylized interdependence structures with $M = 16$ tasks equally assigned to $P=4$ agents for a low and a high level of complexity. The crossed cells indicate inter-dependencies as follows: let $(i,j)$ be coordinates of a crossed cell in row-column order, then performance contribution $\phi(x_{i})$ depends on decision $x_j$.}
%    \label{fig:interactions}
%\end{figure}

The tasks allocated to agents can be similar or distinct. We model this using the pairwise correlations between performance landscapes\footnote{See Verel et al. \cite{verel13} for methodology.}:
\begin{equation}
    \text{corr}(\phi(\mathbf{x}^p_i),\phi(\mathbf{x}^q_i)) = \rho \in [0,1],
\end{equation}
for all $i \in \{1,2,3,4\}$ and $p \neq q$. When $\rho=0$ and $\rho=1$, agents operate on perfectly distinct and perfectly identical performance landscapes, respectively.

\subsection{Agents' compensation}
\label{sec:agents}

Agents $p$'s compensation is based on $p$'s own performance $\phi_{own}$, and the residual performance $\phi_{res}$, defined as the mean of performances of every agent other than $p$:
\begin{equation}
    \phi^p_{res} = \frac{1}{P-1} \cdot \sum_{q \neq p} \phi_{own} (\mathbf{x}^q),
    \label{eq:performance-residual}
\end{equation}
The compensation follows the linear incentive scheme\footnote{In our context linear incentives are as efficient as other contracts inducing non-boundary actions. See \cite[p. 1461]{fischer08}.}: 
\begin{equation}
    \phi_{inc}(\mathbf{x}^p_t) = \alpha \cdot \phi_{own} (\mathbf{x}^p_{t}) + \beta \cdot \phi^p_{res}, 
    \label{eq:utility0}
\end{equation}
where $\alpha + \beta = 1$.

\subsection{Descriptive norms}
\label{sec:social}
In this section we describe our model of descriptive norms\footnote{We implement our version of the Social Cognitive Optimization algorithm. See Xie et al. \cite{xie02} for the original version of the algorithm.}. First of all, we differentiate between two types of tasks, namely \textit{private} and \textit{social} tasks. Private tasks are specific to individual agents and are not subject to descriptive norms, while social tasks concern all agents. For example, in a team, the choice of a server operating system (Debian vs. RHEL) is a private task for the systems administrator, and the choice of a presentation software (\LaTeX vs. PowerPoint) is a social task concerning all team members. In our formulation, private tasks are not relevant to descriptive social norms, while social tasks are. Without loss of generality we use the following convention: the private tasks come first and social tasks come next. For example, if $N_S = 2 $ is the number of social tasks allocated to each agent, then the first 2 tasks are private and the last 2 tasks are social:
\begin{equation}
    \mathbf{x}^p = 
    ( \underbrace{x^p_1,x^p_2}_{\text{private}},
    \underbrace{x^p_{3},x^p_4}_{\text{social}}
    )
\end{equation}

At every time step $t$, agents share their decisions on social tasks with $D=2$ fellow agents, according to the network structure predefined by the modeler.\footnote{We use the bidirectional \textit{ring network}, in which each node is connected to exactly $D=2$ other nodes with reciprocal unidirectional links, where nodes represent agents and the links represent sharing of information.} Every agent stores the shared information in the memory set $L^p$ for up to $T_L=50$ periods, after which the information is ``forgotten'' (removed from $L^p$). Descriptive norms are not formed until time period $T_L$.

The extent to which agent $p$'s decisions $\mathbf{x}^{p}_{t}$ comply with the descriptive social norms is computed as the average of the matching social bits in the memory:
\begin{equation}
    \phi_{soc} (\mathbf{x}^p_{t}) = \begin{cases}
        \displaystyle\frac{1}{|L^p_t|} \sum_{\mathbf{x^L} \in L^p_t} \frac{[x^p_3==x^L_3]+[x^p_4==x^L_4]}{2}, & t > T_L \\
         0, & t \leq T_L
    \end{cases}
    \label{eq:norms}
\end{equation}
where $|L^p_t|$ is the number of entries in agent $p$'s memory at time $t$, and the statement inside the square brackets is equal to $1$ if true, and $0$ if false \cite{iversion62}.

%Depending on the correlation among the tasks assigned to different employees, the compliance with the social norms can help or hurt the organizational performance. Section \ref{sec:multiobj} describes how the agents handle the multiple objectives in detail.

\subsection{Search process}
\label{sec:discovering}

At time $t$, agents can observe their own performance in the last period, $\phi_{own} (\mathbf{x}^p_{t-1})$, and the decisions of all team members in the last period \textit{after} they are implemented, $\mathbf{x}_{t-1}$.

In order to come up with new solutions to their decision problems, agents perform a search in the neighbourhood of $\mathbf{x}_{t-1}$ as follows: agent $p$ randomly switches one decision $x_i \in \mathbf{x}^p$ (from $0$ to $1$, or vice versa), and assumes that other agents will not switch their decisions\footnote{Levinthal \cite{levinthal97} describes situations in which agents switch more than one decision at a time as \textit{long jumps} and states that such scenarios are less likely to occur, as it is hard or risky to change multiple processes simultaneously.}. We denote this vector with one switched element by $\hat{\mathbf{x}}^p_t$. 

Next, the agent has to make a decision whether to stick with the status quo,  $\mathbf{x}^p_t$, or to switch to the newly discovered $\hat{\mathbf{x}}^p_t$. The rule for this decision is to maximize the weighted sum of the performance-based compensation and the compliance with the descriptive social norms:

\begin{equation}
    \mathbf{x}^p_t = \underset{\mathbf{x} \in \{\mathbf{x}^p_{t-1}, \hat{\mathbf{x}}^p_{t}\}}{\arg\max}   w_{1} \cdot \phi_{inc} (\mathbf{x}) + w_{2} \cdot \phi_{soc} (\mathbf{x}),
    \label{eq:goalprog}
\end{equation}
where $w_1 + w_2 = 1$.

\subsection{Process overview, scheduling and main parameters}
\label{sec:process}
\begin{figure}[!t]
	\centering
	\includegraphics[width=0.93\linewidth]{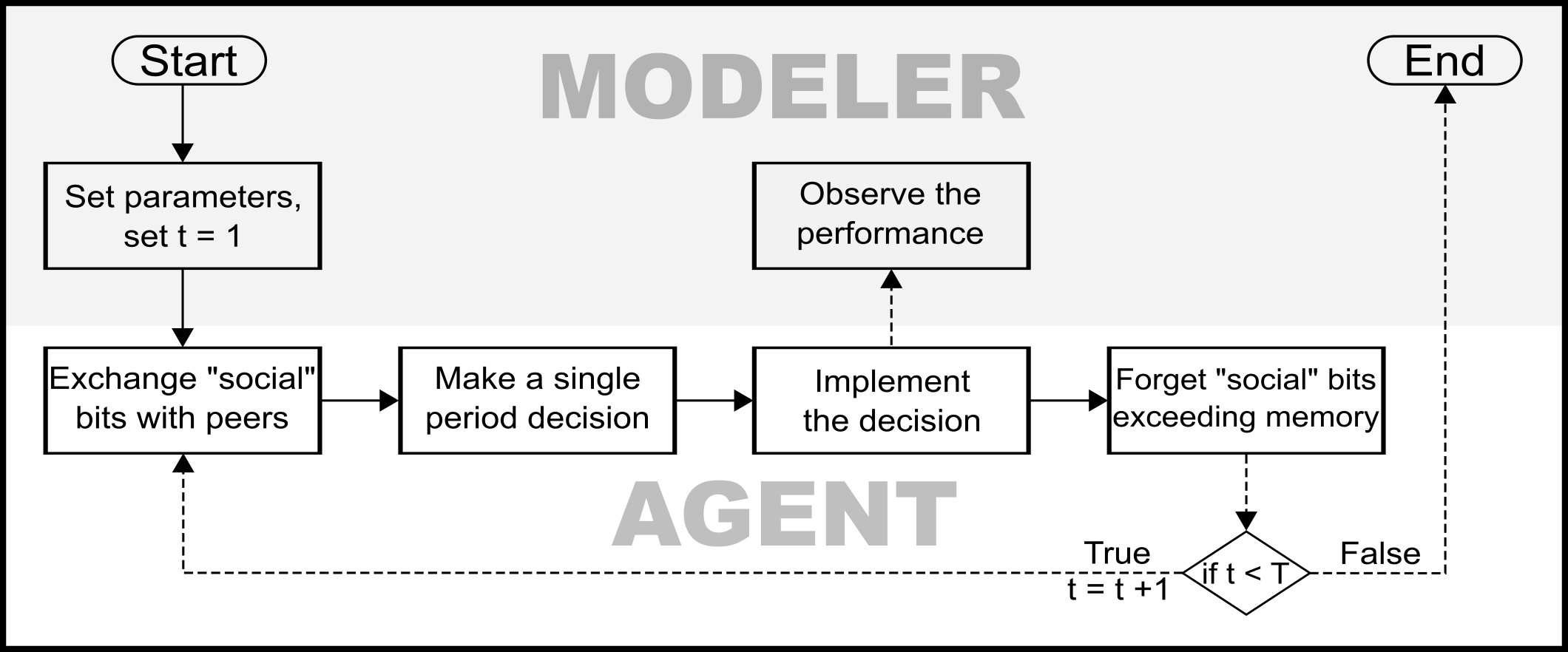}
	\caption{Process overview}
	\label{fig:funcflowchart}
	\vspace*{-2em}
\end{figure}

The simulation model has been implemented in \textit{Python 3.8} and \textit{Numba} just-in-time compiler. Every simulation round starts with the initialization of the agents' performance landscapes, the assignment of tasks to $P=4$ agents\footnote{For reliable results, we generate the entire landscapes before the simulation, which is feasible for $P=4$ given modern computing limitations. Our sensitivity analyses with simpler models without entire landscapes suggest that the results also hold for $P=5,6,7$.}, and the initialization of an $M=16$ dimensional bitstring as a starting point of the simulation run. After initialization, agents perform the \textit{hill climbing} search procedure outlined above and share information regarding their social decisions in their social networks. The observation period $T$, the memory span of the employees $T_L$, and the number of repetitions in a simulation, $R$, are exogenous parameters, whereby the latter is fixed on the basis of the coefficient of variation. Figure \ref{fig:funcflowchart} provides an overview of this process and Tab. \ref{tab:params} summarizes the main parameters used in this paper.

\section{Results}
\label{sec:results}
We perform $R=1000$ simulations for every combination of parameters presented in Tab. \ref{tab:params}. Our main variable of interest is the dynamics of teams' normalized overall performance, $\Phi^r_t$, for all $r \in \{1,2,...,R\}$ and $t \in \{1,2,...,T\}$.\footnote{As the performance landscapes are randomly generated, we normalize the team performance by the maximum performance attainable in the current simulation run $r$ to ensure comparability}. We denote the normalized performance at time $t$, averaged over all simulation runs by:
\begin{equation}
    \overline{\Phi_t} = \frac{1}{R} \sum_{r=1}^R \Phi^r_t
\end{equation}

\begin{figure}[!tb]
	\centering
	\begin{minipage}{0.04\linewidth}
		\rotatebox{90}{\parbox{9em}{\centering $\alpha = 1.0$\\Overall performance}}
	\end{minipage}
	\begin{minipage}{0.47\linewidth}
		\centering
	    {Low complexity}	    
		\includegraphics[width=\linewidth]{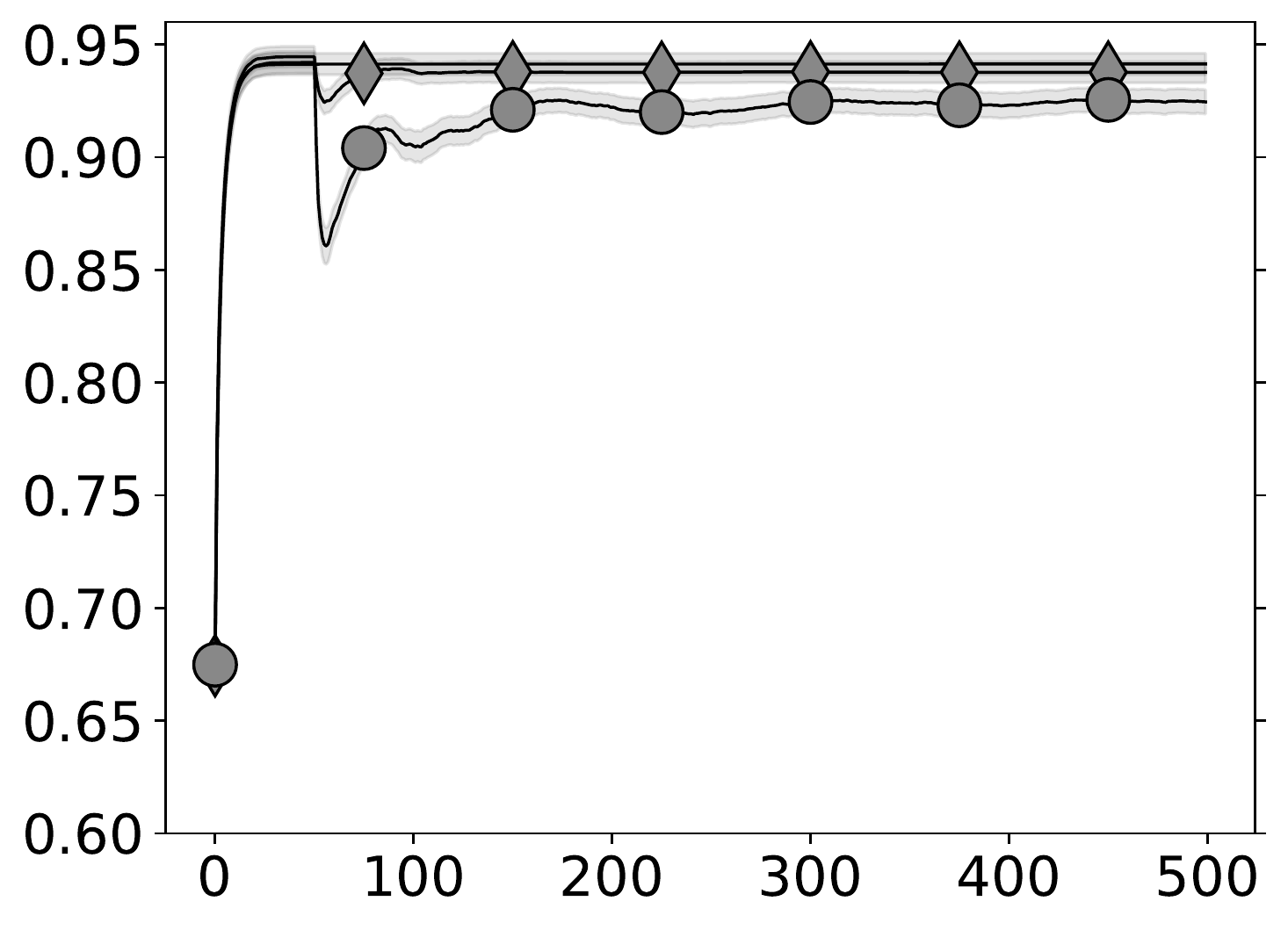}
	\end{minipage}
	\begin{minipage}{0.47\linewidth}
		\centering
	    {High complexity}
		\includegraphics[width=\linewidth]{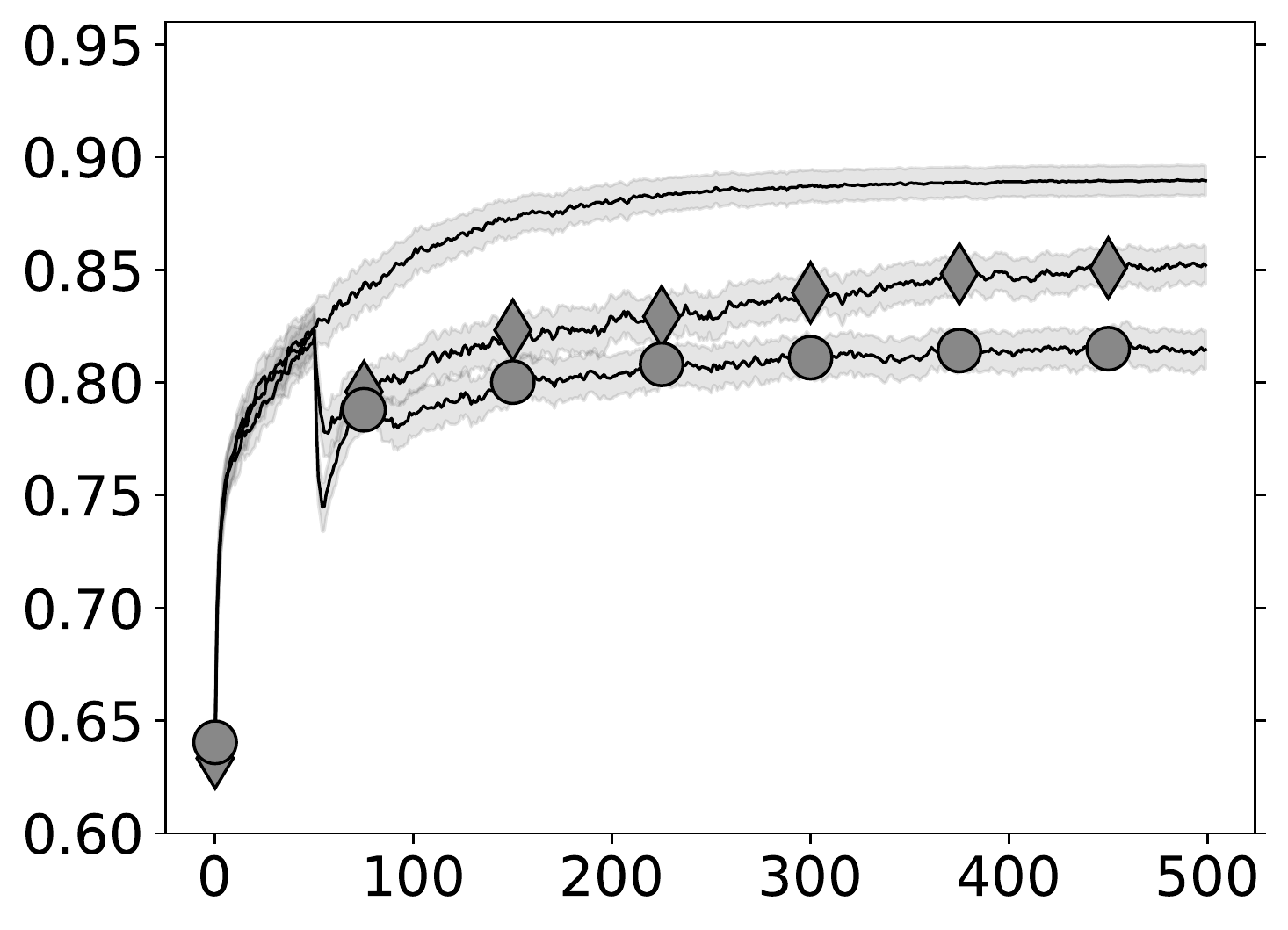}
	\end{minipage}
	
	\begin{minipage}{0.04\linewidth}
		\rotatebox{90}{\parbox{9em}{\centering $\alpha = 0.75$\\Overall performance}}
	\end{minipage}
	\begin{minipage}{0.47\linewidth}
		\includegraphics[width=\linewidth]{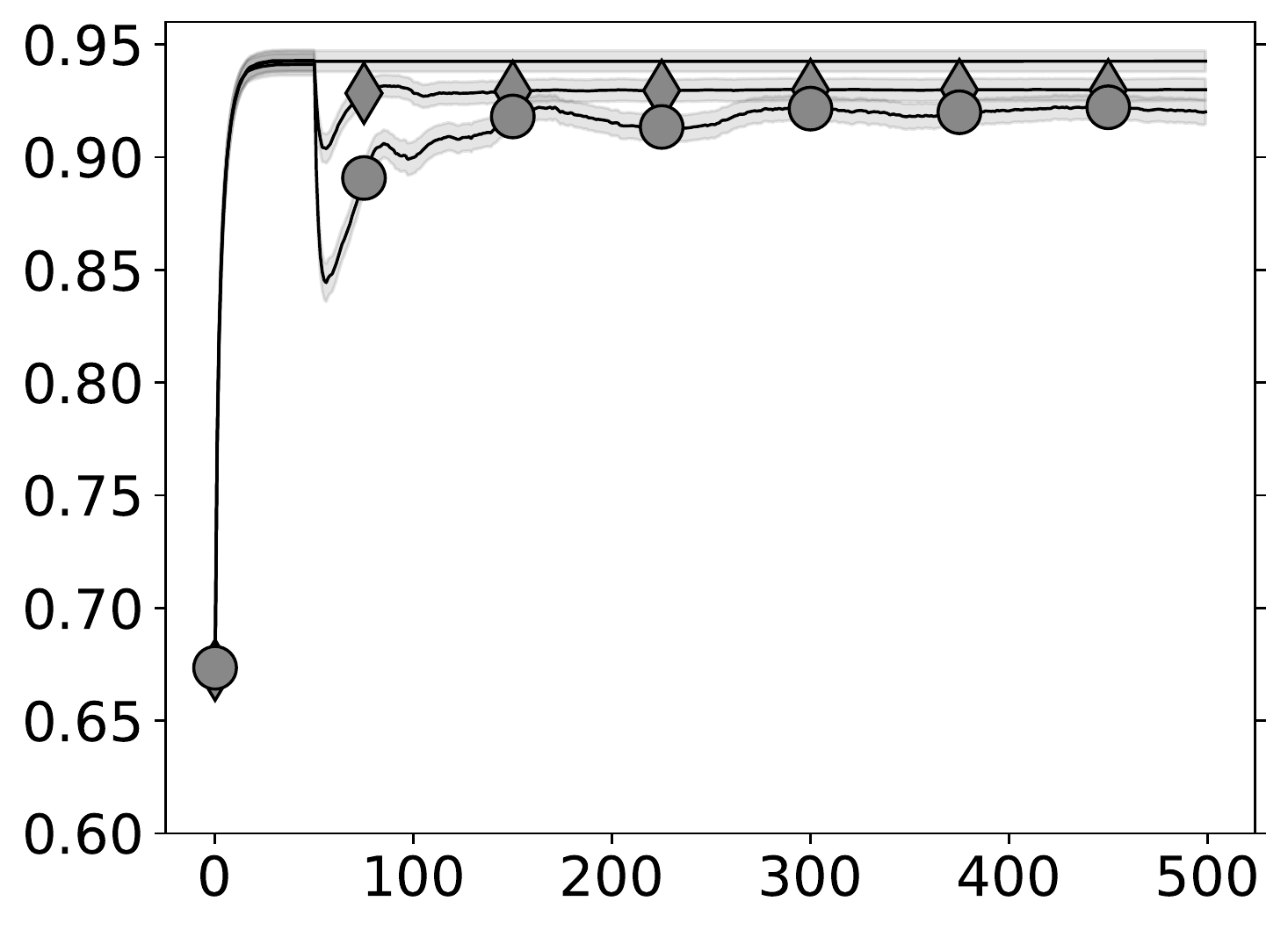}
	\end{minipage}
	\begin{minipage}{0.47\linewidth}
		\includegraphics[width=\linewidth]{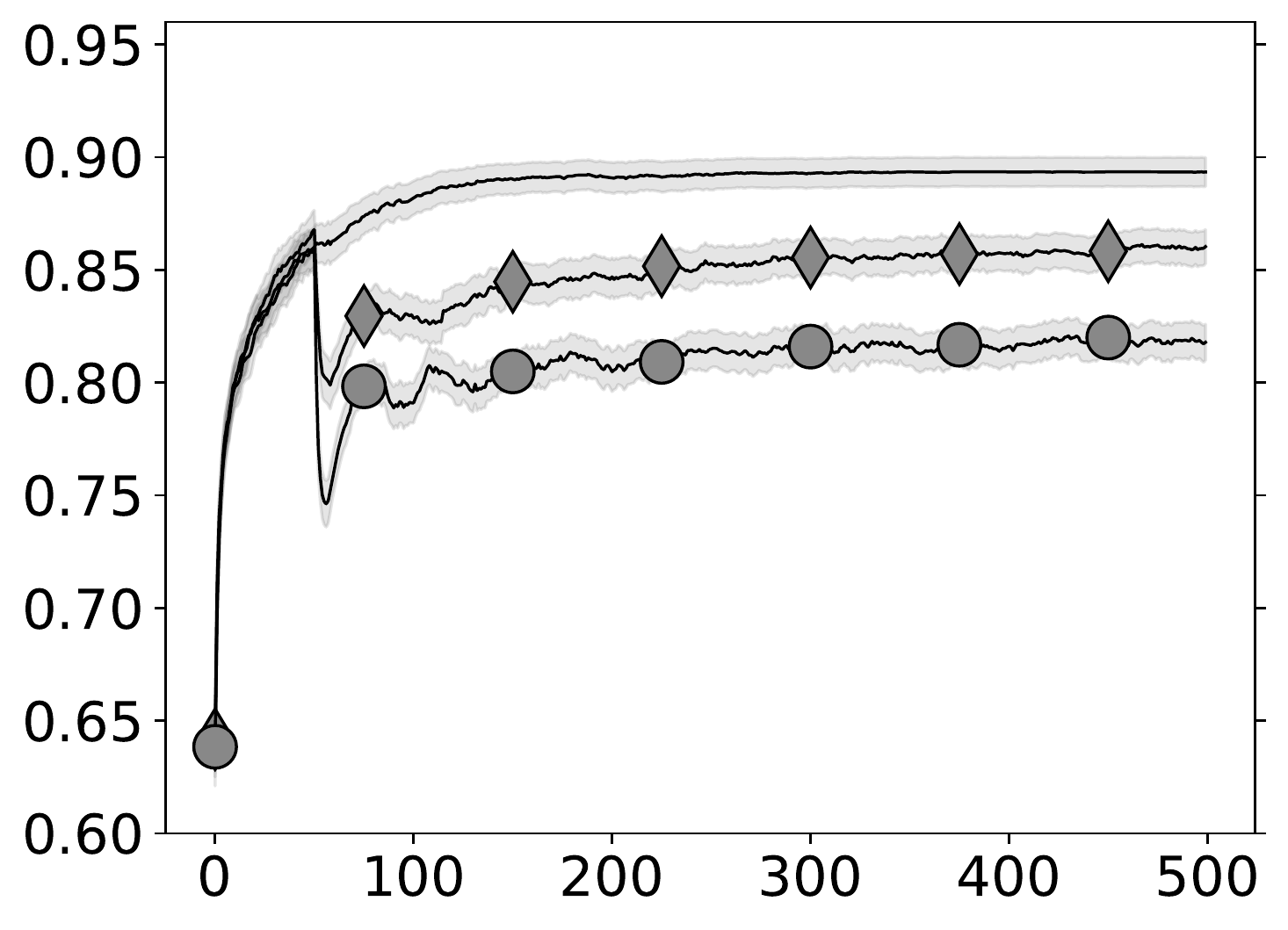}
	\end{minipage}
	
	\begin{minipage}{0.04\linewidth}
		\rotatebox{90}{\parbox{9em}{\centering $\alpha = 0.25$\\Overall performance}}
	\end{minipage}
	\begin{minipage}{0.47\linewidth}
		\centering
		\includegraphics[width=\linewidth]{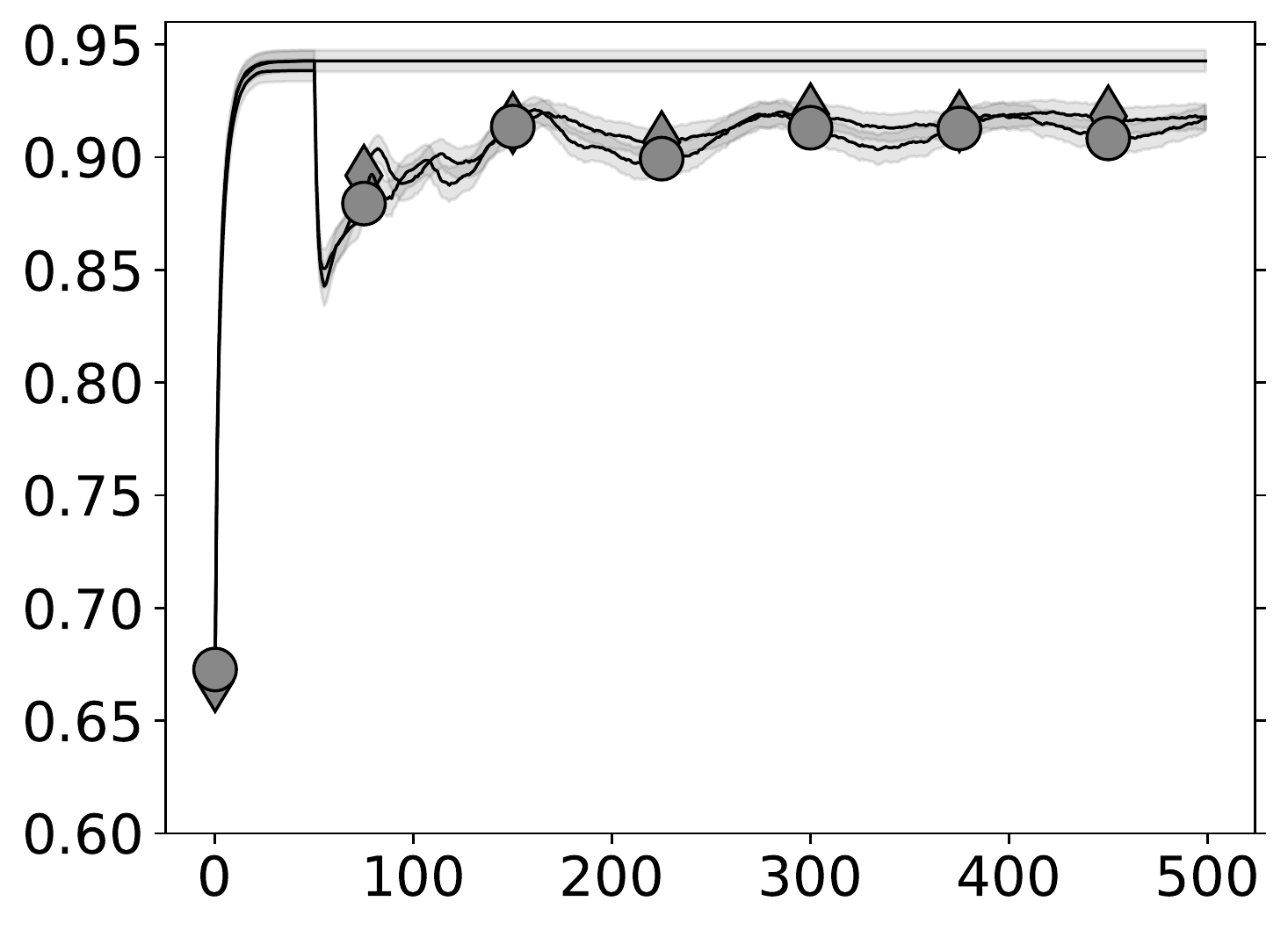}
		{Time steps}
	\end{minipage}
	\begin{minipage}{0.47\linewidth}
		\centering
		\includegraphics[width=\linewidth]{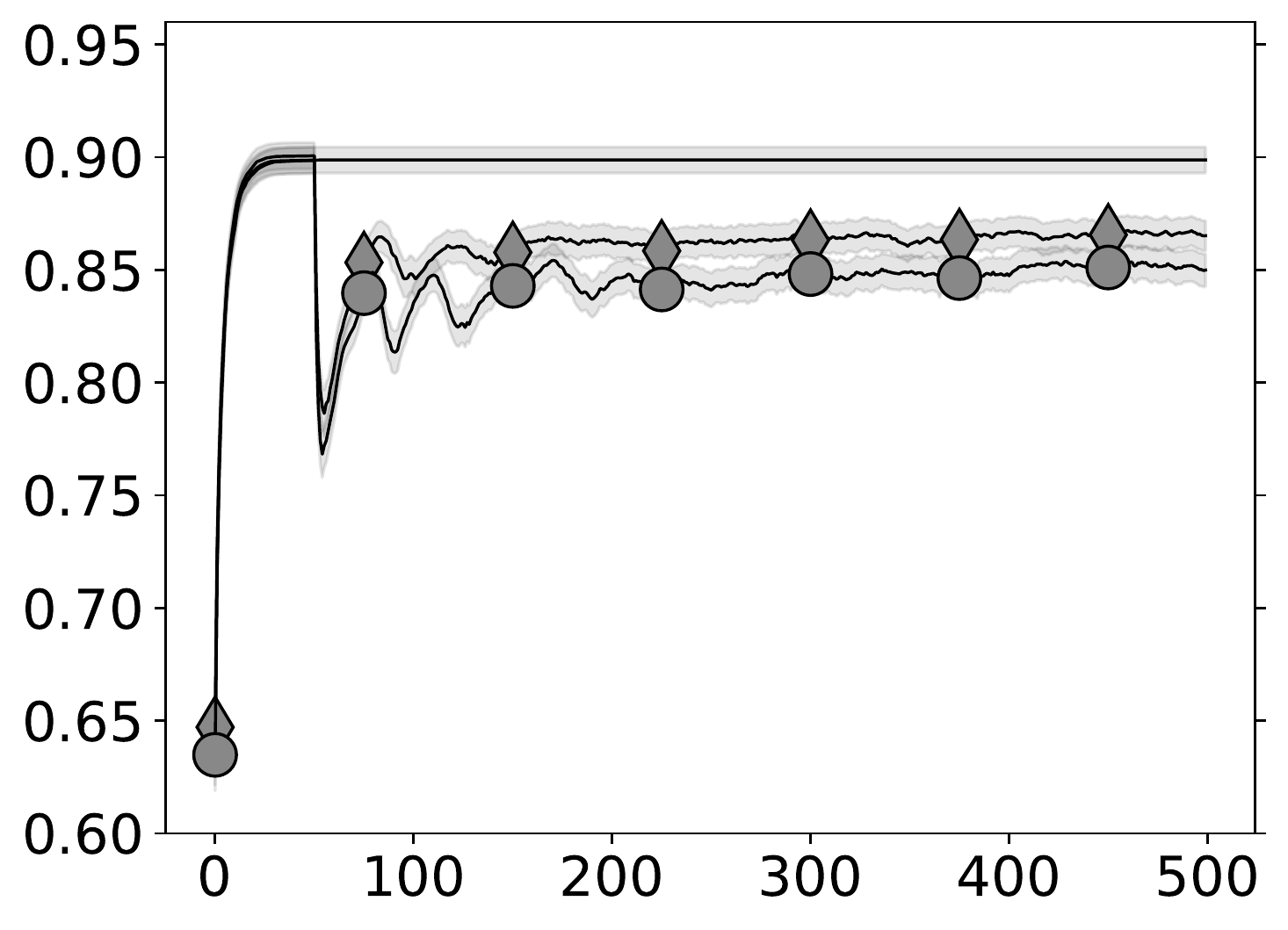}
		{Time steps}
	\end{minipage}
	
	\includegraphics[width=\linewidth]{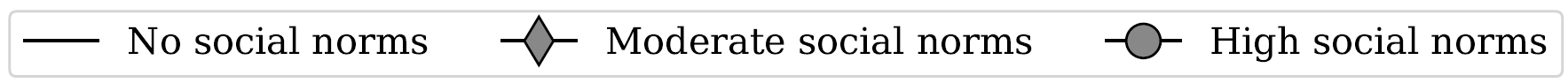}
	
	\caption{Main results based on parameters defined in Tab. \ref{tab:params}. The plots show normalized performance for $T=500$ time steps, averaged over $R=1000$ simulation runs with $99.9\%$ confidence interval.}
	\label{fig:results}
	\vspace*{-2em}
\end{figure}

The main results of the simulation for parameters defined in Tab. \ref{tab:params} are presented in Fig. \ref{fig:results}. The figure contains 8 sub-figures for 2 levels of complexity described in Eq. \ref{eq:payoff}, and 3 incentive schemes with different weights of team performance described in Eq. \ref{eq:utility0}. Each sub-figure includes 3 time series of normalized overall performance $\overline{\Phi_t}$ over $T=500$ periods for different weights of social norms in agents' decision rules, defined in Eq. \ref{eq:goalprog}.

\begin{table}[!t]
    \centering
    \caption{Main parameters}
    \label{tab:params}
    \begin{tabular}{c p{18em} >{\raggedright\arraybackslash}p{8em}}
        Parameter & Description                                 & Value         \\
        \hline
        $M$       & Total number of tasks                       &   16 \\
        $P$       & Number of agents                            & 4             \\
        $N$       & Number of tasks assigned to a single agent  & 4             \\
        $[K,C,S]$ & Internal and external couplings             & $[2,0,0]$, $[2,2,2]$    \\                  
        $\rho$    & Pairwise correlation between landscapes  & 0.3      \\
        $T_L$     & Memory span of agents                       & 50            \\
        $N_{S}$   & Number of social tasks                      & 2    \\
        $D$       & Number of connected peers (node degree)     & 1             \\
        $T$       & Observation period                          & 500           \\
        $R$       & Number of simulation runs per scenario      & 1000           \\
    $[ w_1, w_2]$ & Weights for incentives $\phi_{inc}$ and compliance with the social norms $\phi_{soc}$                                                                                 & $[1,0]$, $[0.7,0.3]$, $[0.5,0.5]$ \\
$[\alpha,\beta]$ & Shares of own and residual performances      & $[1,0]$, $[0.75,0.25]$, $[0.25,0.75]$ \\
        \hline
    \end{tabular}
    \vspace*{-2em}
\end{table}

First of all, we observe that the performance is lower at all periods in environments with high complexity than in environments with low complexity (i.e. all values in the left column are greater than their counterparts in the right column), which is in line with the previous research \cite{kauffman91,levinthal97}.

Second of all, we observe in all sub-figures that, as agents put more emphasis on social norms in their decision rules, the team's overall performance drops, which is more pronounced for environments with high complexity.

Third, we observe that in the absence of social norms (top curves in all sub-plots) the incentive schemes  do not have a significant effect for tasks with low complexity, while incentive schemes that put more emphasis on team performance decrease the time it takes to converge for tasks with high complexity (i.e. the curves skew left as we move from upper to lower sub-plots in the right column, and do not change in the left column). This finding is in line with the existing literature \cite{rivkin00}.

Fourth, we find that in presence of social norms the incentive schemes do not have a significant effect on the long-run team performance most of the time, with two exceptions: (i) if the task has a low complexity and agents put a moderate emphasis on social norms, the long-run performance increases for incentive schemes that put less weight on team performance (i.e. curve denoted by \tikz\node[rectangle,draw,xscale=0.7,rotate=45,fill=gray] {}; in the left column shifts upward as we move from lower to upper sub-plots) and (ii) if the task has a high complexity and agents put a high emphasis on social norms, the long-run performance increases for incentive schemes that put more weight on team performance (i.e. curve denoted by \tikz\node[circle,draw,yscale=0.9,xscale=0.9,fill=gray] {}; in the right column shifts upward as we move from upper to lower sub-plots).

Next, we perform sensitivity analyses on variables that are essential to our formulation of social norms: the fixed network degree $D$, the number of social tasks $N_S$, and the correlation between landscapes $\rho$. For the sensitivity analyses, we consider the baseline case with no team-based incentives ($\alpha=1$), high social norms ($w_1 = w_2 = 0.5$). All other parameters are taken from Tab. \ref{tab:params}. Moreover, we also include the case with no social norms ($w_1 = 1, w_2 = 0$) as a benchmark.

\begin{figure}[!t]
    \centering
    \begin{minipage}{0.02\linewidth}
        \rotatebox[]{90}{\hspace*{2em}Overall performance}
    \end{minipage}
    \begin{minipage}{0.475\linewidth}
    	\centering
        {Low complexity}
        \includegraphics[width=\linewidth]{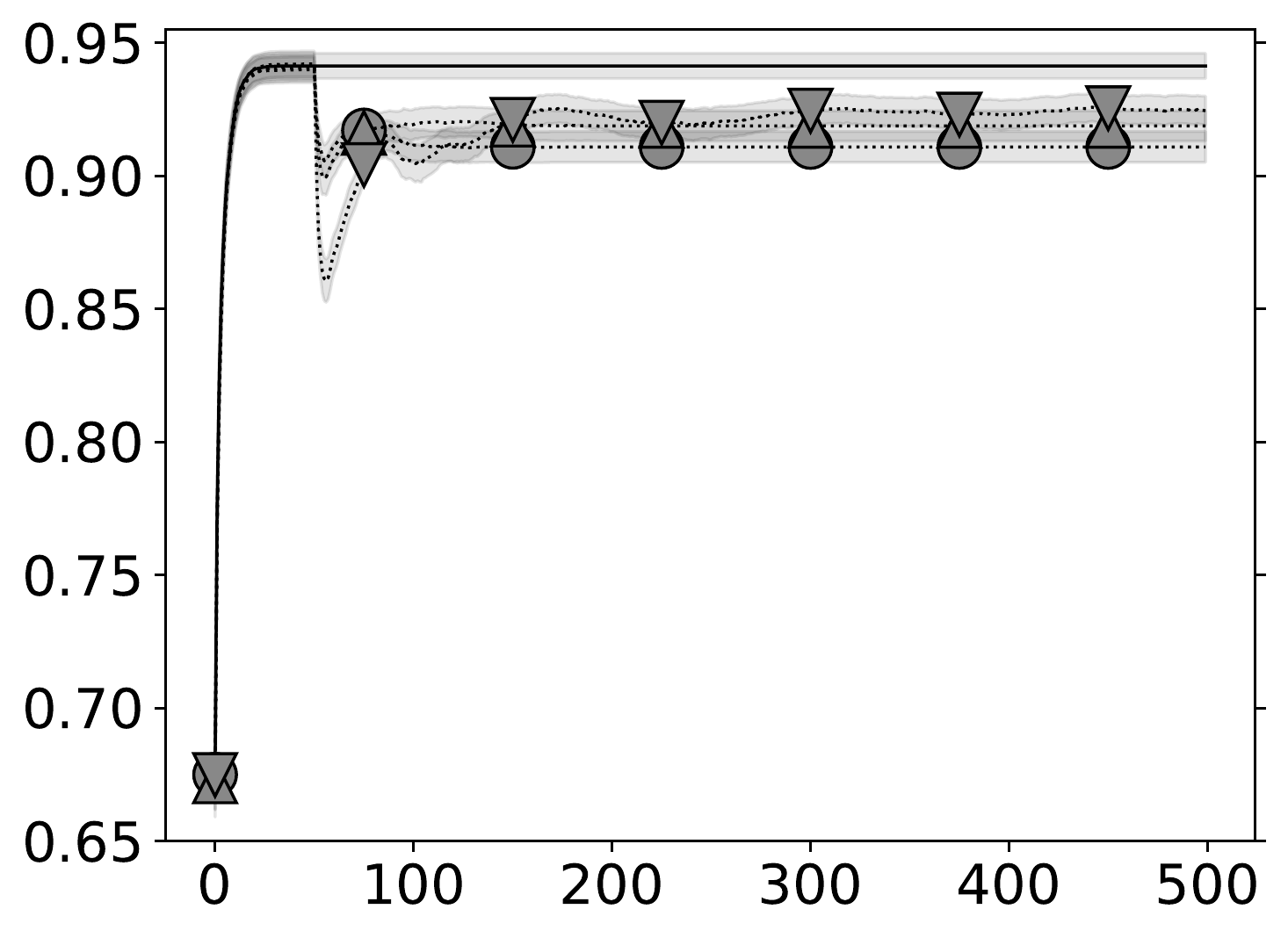}
        {Time steps}
    \end{minipage}
    \begin{minipage}{0.475\linewidth}
    	\centering
        {High complexity}
        \includegraphics[width=\linewidth]{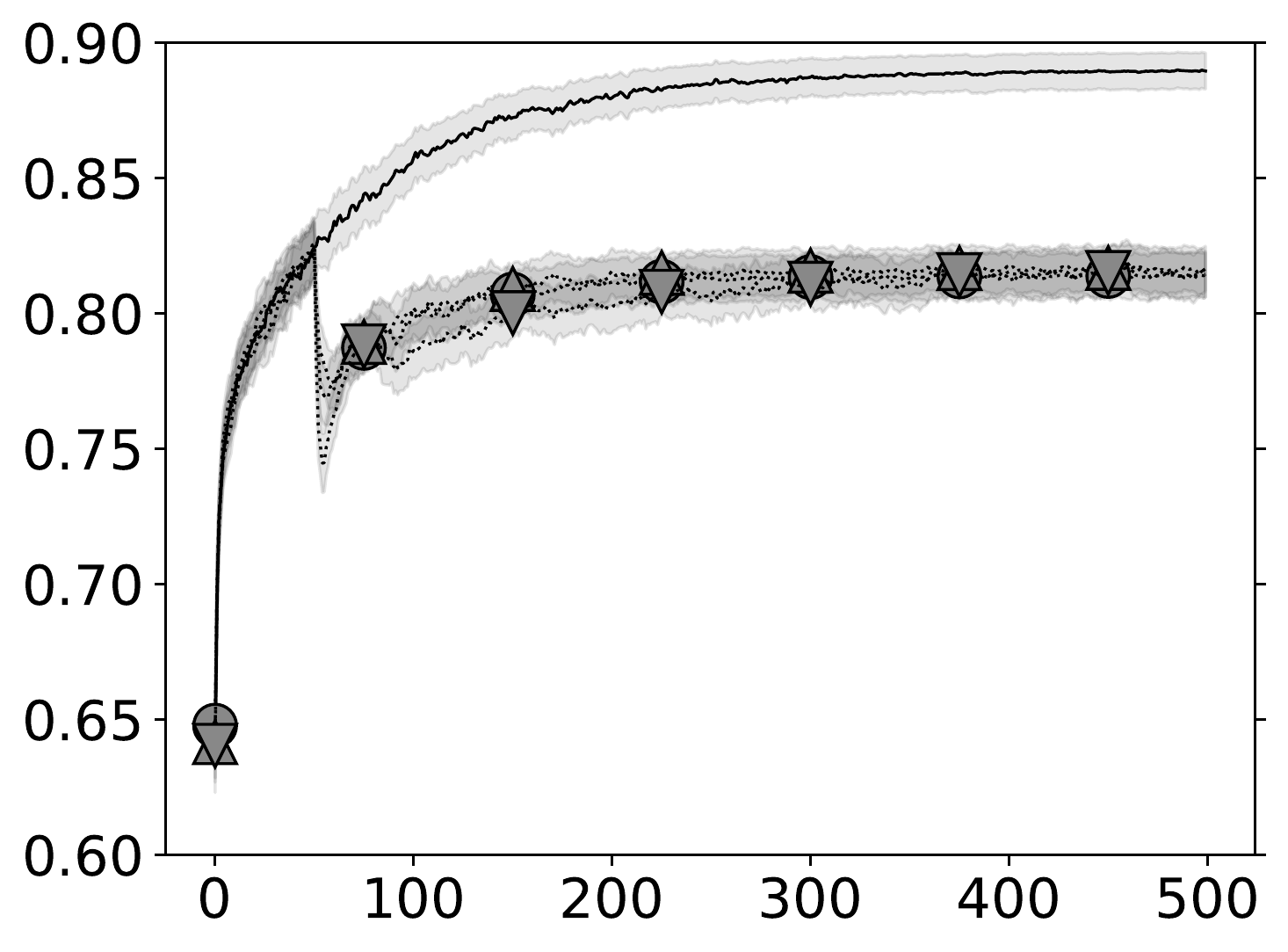}
        {Time steps}
    \end{minipage}
    
    \includegraphics[width=0.67\linewidth]{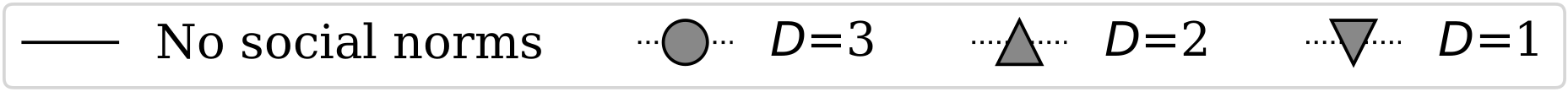}
    
    \caption{Overall performance for different values of $D$ ($99.9\%$ confidence interval)}
    \label{fig:sensdeg}
    \vspace*{-1.5em}
\end{figure}

\begin{figure}[!b]
    \vspace*{-2em}
    \centering
    \begin{minipage}{0.02\linewidth}
        \rotatebox[]{90}{\hspace*{2em}Overall performance}
    \end{minipage}
    \begin{minipage}{0.475\linewidth}
    	\centering
        {Low complexity}
        \includegraphics[width=\linewidth]{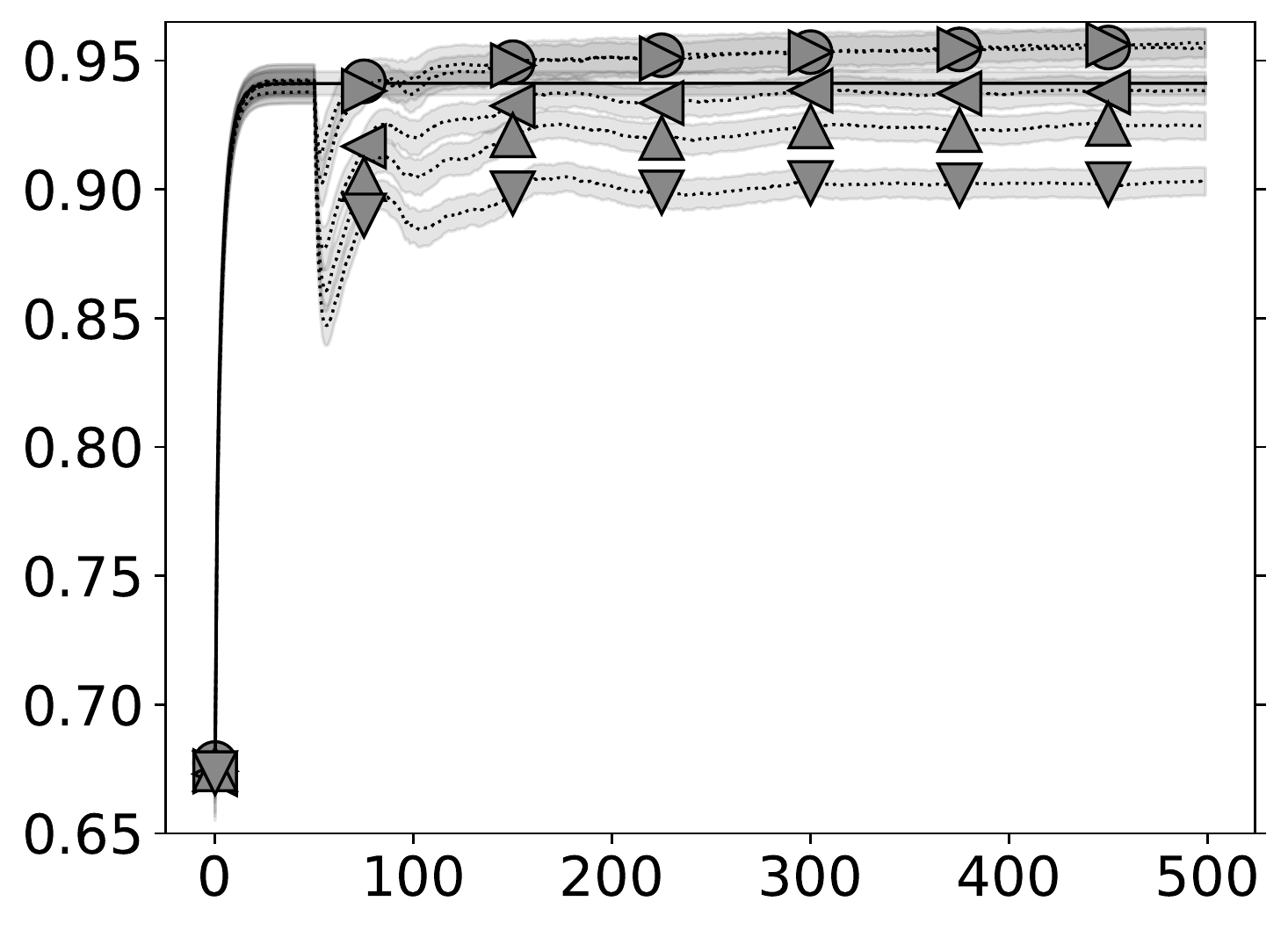}
        {Time steps}
    \end{minipage}
    \begin{minipage}{0.475\linewidth}
        \centering
        {High complexity}
        \includegraphics[width=\linewidth]{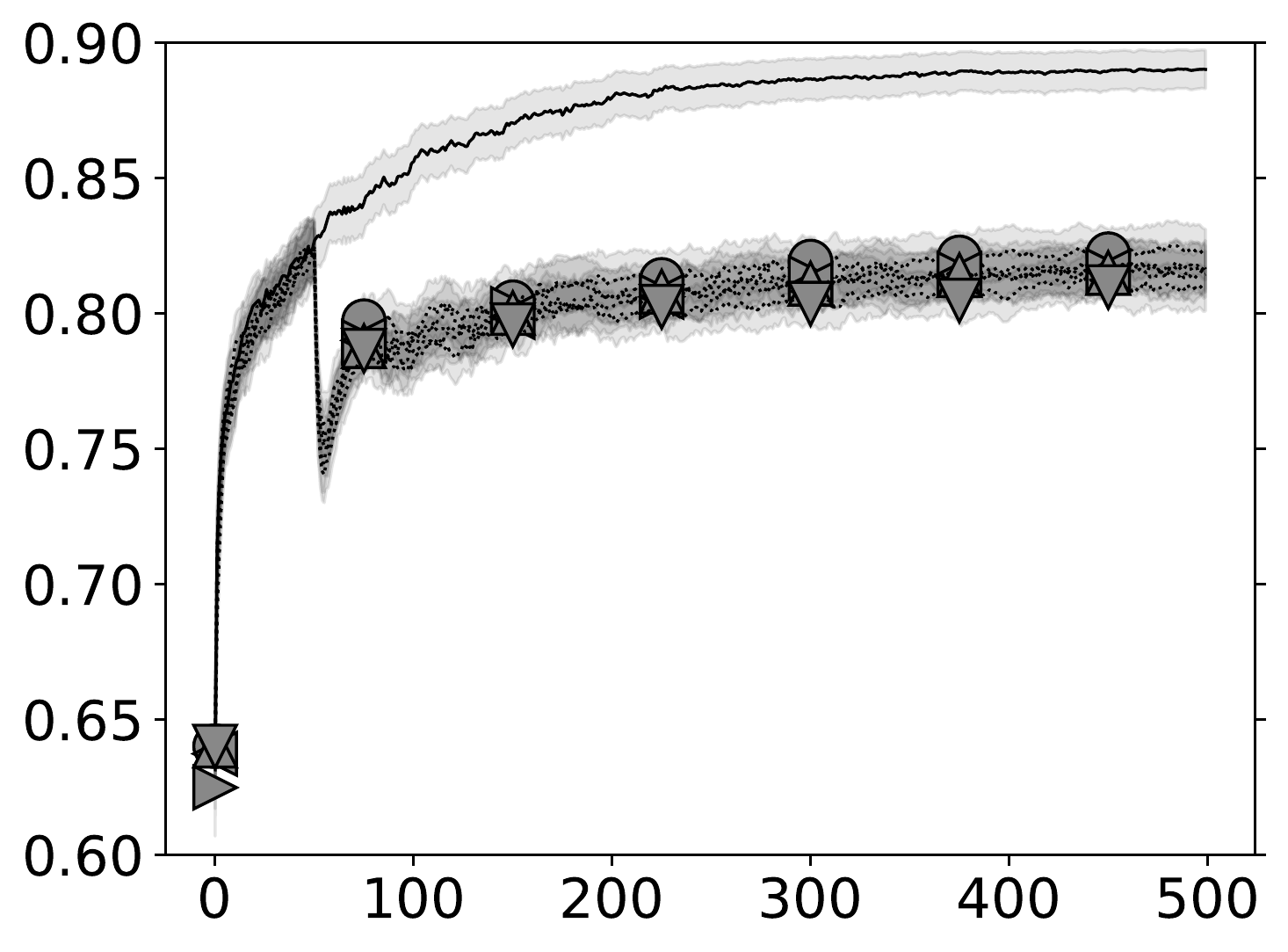} 
        {Time steps}
    \end{minipage}
    
    \includegraphics[width=\linewidth]{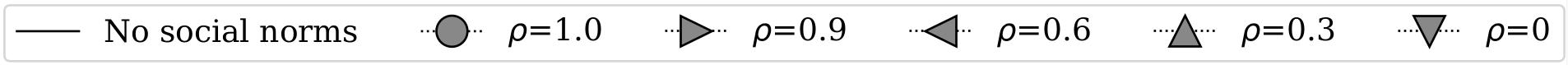}
    \caption{Overall performance for different values of $\rho$ ($99.9\%$ confidence interval)}
    \label{fig:sensrho}
    \vspace*{-1em}
\end{figure}

\begin{figure}[!t]
    \centering
    \begin{minipage}{0.02\linewidth}
        \rotatebox[]{90}{\hspace*{2em}Overall performance}
    \end{minipage}
    \begin{minipage}{0.475\linewidth}
        \centering
        {Low complexity}
        \includegraphics[width=\linewidth]{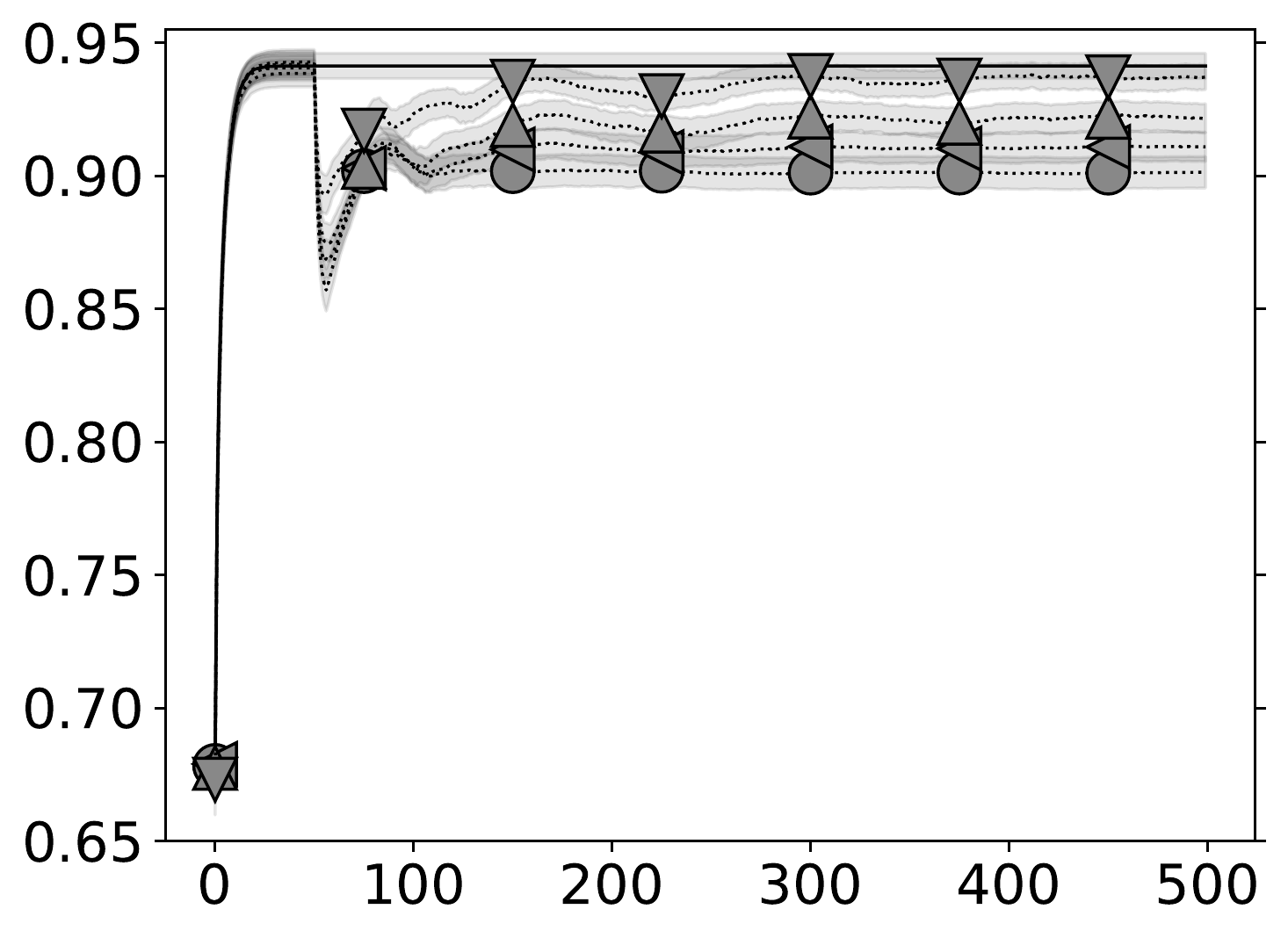}
        {Time steps}
    \end{minipage}
    \begin{minipage}{0.475\linewidth}
        \centering
        {High complexity}
        \includegraphics[width=\linewidth]{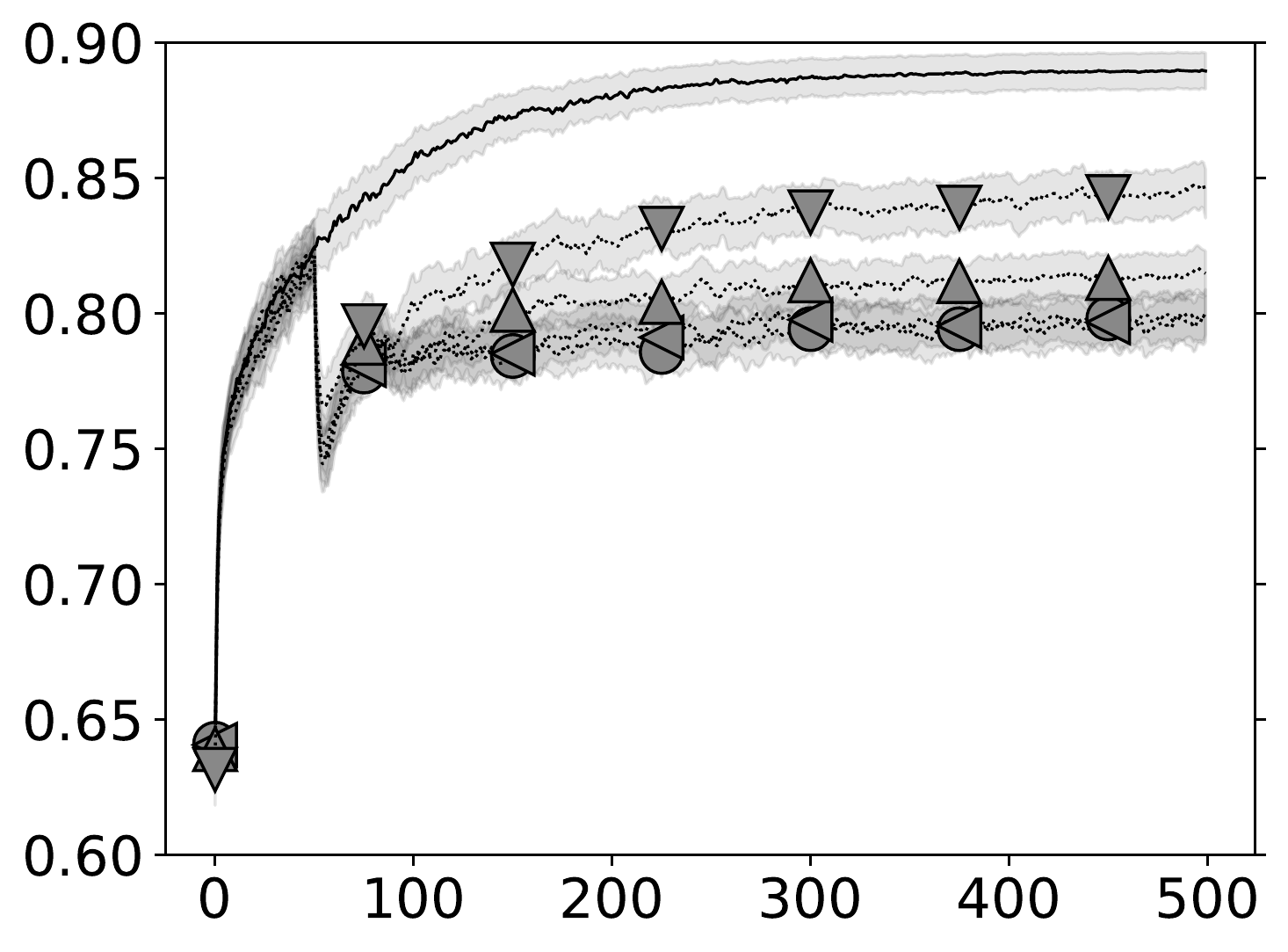}
        {Time steps}
    \end{minipage}
    
    \includegraphics[width=0.83\linewidth]{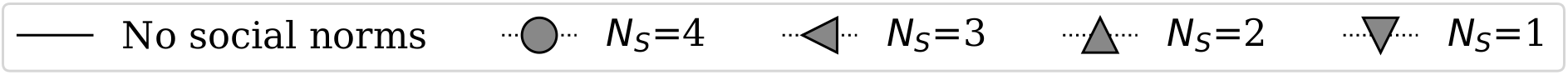}
    
    \caption{Overall performance for different values of $N_S$ ($99.9\%$ confidence interval)}
    \label{fig:sensnsoc}
    \vspace*{-1.5em}
\end{figure}

Fig. \ref{fig:sensdeg} illustrates that network degree $D$ does not sıgnıfıcantly change the effect of socıal norms on the overall performance. Fig. \ref{fig:sensnsoc} shows that as the number $N_S$ of social bits increases, the effect of descriptive social norms gets more pronounced. Fig. \ref{fig:sensrho} shows that the correlation between landscapes does not significantly affect the overall performance in environments with high complexity. However, in environments with low complexity, not only does the positive correlation offset the performance drop caused by the descriptive social norms, but it also may improve the performance for $\rho \geq 0.9$.

\section{Conclusion}
\label{sec:conclusion}
A thorough examination of alternatives is costly for individuals and for this reason they turn to descriptive social norms to avoid computational costs and use them as a shortcut by complying with the perceived consensus of their colleagues. We have provided evidence that this comes at a cost for the overall performance of a team, unless tasks are highly correlated. We have analyzed the effect for different environments and levels of communication and found that while the number of common tasks increases the effect of the social norms on performance, the level of communication does not. Finally, we have confirmed that the established finding that team-based incentives increase overall performance for highly complex tasks also works in presence of social norms.

\bibliographystyle{splncs04}
\bibliography{refs.bib}
\end{document}